\documentclass[twocolumn,aps,pre,showpacs]{revtex4}
\usepackage[latin1]{inputenc}
\usepackage{latexsym}
\usepackage{amssymb}
\usepackage{amsmath}
\usepackage{graphicx}
\begin{document}
\def\d{{\rm d}}
\def\ex{{\rm e}}
\def\i{{\rm i}}
\def\rv{r_{\rm v}}
\def\e{{\bf e}}
\def\a{{\bf a}}
\def\v{{\bf v}}
\def\u{{\bf u}}
\def\x{{\bf x}}
\def\r{{\bf r}}
\def\k{{\bf k}}
\def\h{{\bf h}}
\def\y{{\bf y}}
\def\smalze{{\scriptscriptstyle (0)}}
\def\smalun{{\scriptscriptstyle (1)}}
\def\smaldu{{\scriptscriptstyle (2)}}
\def\smalk{{\scriptscriptstyle (k)}}
\def\bxi{{\boldsymbol{\xi}}}
\def\bnu{{\boldsymbol{\nu}}}
\def\beq{\begin{equation}}
\def\eeq{\end{equation}}
\title{
Clustering and collision of inertial particles in random velocity fields
}
\author{Piero Olla}
\affiliation{
ISAC-CNR, and INFN Sez. Cagliari, I--09042 Monserrato, Italy.
}
\date{\today}

\begin{abstract}
The influence of clustering on the collision rate
of inertial particles in a smooth random velocity field,
mimicking the smaller scales of a turbulent flow, is 
analyzed.
For small values of the the ratio between
the relaxation time of the particle velocity and the 
characteristic time of the field, the effect of
clusters is to make more energetic collisions less likely.
The result is independent of the flow dimensionality
and  is due only to the origin of collisions in the process
of caustic formation. 

\end{abstract}

 \maketitle

The transport of finite size particles in turbulent flows is a common occurrence 
in several environments:  raindrops in clouds \cite{shaw03}, 
plankton in oceans \cite{ruiz04}, sprays in industrial flows 
\cite{crowe98}, to make some examples. Due to inertia, these particles undergo 
clustering phenomena that have been observed in numerical simulations 
\cite{cencini06} ,
experiments \cite{kostinski01}, and have been the subject 
of substantial theoretical
study \cite{wang93,teorici,elperin96,balkovsky01,duncan05}.

Although contributing to particle segregation \cite{brooke94}, 
spatial inhomogeneities in the turbulence do not appear to be an essential 
factor. 
What seems to be important is the ability of the particles to catch
one another in their motion, as they slip with respect to the fluid, 
a circumstance that is most evident in one dimension (1D)
\cite{deutsch85,wilkinson03}. In more than 1D, 
the process is more
complicated and 
an important role is  played by the preferential 
concentration of heavy (light) particles in the strain (vortical) regions
of the flow \cite{wang93}. 

An important motivation for the interest in clustering is the 
contribution to particle collision and coalescence, and 
it has been suggested that this is an important ingredient in the 
process of rain formation \cite{sundaram97,shaw03}. 
What is observed is the simultaneous onset of 
concentration fluctuations and increased collision rates, when
the relaxation time of the particle velocity relative to the fluid
becomes of the  order of the turnover time of the fastest turbulent
eddies \cite{sundaram97}. 
For sufficiently small (and sufficiently dense) spherical particles,
this relaxation time is the Stokes time
$\tau_S=1/18\ a^2\lambda/\nu_0$, where $a$ is the particle diameter,
$\lambda$ is the ratio of the particle to fluid density and $\nu_0$ is
the kinematic viscosity of the fluid \cite{maxey83}.

When inertia is sufficiently high, the so called 
sling effect ensues  \cite{falkovich02}:
particle and fluid trajectories detach on the
scale of eddies with turnover time $\tau_S$, and their 
velocity will determine the particle collision velocity.

The collision rate $R_{coll}$ depends both on the particle concentration 
$n(\x,t)$ and on the relative velocity $\bnu$ in the particle pairs; 
for binary collisions:
\beq
R_{coll}\sim a^2\,\langle n(a)n(0)\rangle\,\langle \nu|r=a\rangle,
\label{rate}
\eeq
with $\r$ the particle separation.
(Similar expressions will hold for the coalescence rate,
with a more complicate function of $\bnu$
in the conditional average). 
Thus, both the sling effect and clustering, through the factors
$\langle \nu|\r\rangle$ and $\langle n(a)n(0)\rangle$, respectively,
may be expected to enhance collisions. 
It was suggested in \cite{falkovich02,wilkinson06}, however,
that the dominant contribution to collision may be the sling effect
and there is even some indication \cite{olla07} that clustering 
may hinder rather than enhance collisions.

The problem of how clustering and the dynamics of
the two-particle velocity distribution in general 
influence each other, is far from trivial, with 
caustic formation potentially playing an important role
\cite{falkovich02,wilkinson05}.
Purpose of this rapid communication is to understand whether it
is possible to identify a clustering contribution to the velocity 
dynamics, and if this is associated with collision enhancement or 
hindering.

Let us consider the dynamics of an inertial particle suspension in a smooth, 
incompressible random velocity field $\u(\x,t)$, with correlation 
time $\tau_E$, variance $\langle u_\alpha^2\rangle=\sigma^2_u$, $\alpha=1,2,3$
and correlation length $\rv\sim\sigma_u\tau_E$.  Particles in a turbulent
flow, with $\tau_S$ 
shorter than the Kolmogorov time, here identified with $\tau_E$,
will see turbulence precisely in this way, and this is 
appropriate for most aerosols of atmospheric interest 
\cite{shaw03,falkovich02}. Conversely,  
the opposite regime $\tau_S>\tau_E$ could be interpreted as a model for the effect of
eddies at scale $\rv$ on particles with $\tau_S$ corresponding to 
the turnover time of larger eddies. This limit is more
relevant for industrial flows than for rain formation, as
the relative motion of larger droplets in clouds
is dominated by 
the 
different gravitational settling velocities 
of droplets of different size
\cite{shaw03}.

The two regimes of large and small Stokes number $S=\tau_S/\tau_E$
are qualitatively differents, with the clustering maximum occurring
somewhere at $S\lesssim 1$ \cite{wang93,fessler94}. For small $S$,
the particle
phase is monodisperse in velocity over most of the fluid volume 
\cite{elperin96,boffetta07}. The sling effect,
due to the spatial correlation of $\u$, occurs in coherent way 
through the formation of caustics, i.e. regions 
of crossing of particle jets with different velocity 
\cite{falkovich02}.
The two-particle 
dynamics, relevant for the description of clustering and binary collision,
is described by the equation for the velocity difference
$\tau_S\dot\bnu+\bnu=\Delta_\r\u$, $\Delta_\r\u(\x,t)
\equiv\u(\x+\r,t)-\u(\x,t)$, and, for $S\ll 1$, 
$\langle\nu^2|r\rangle\sim \langle |\Delta_\r\u|^2\rangle\to 0$
for $r\to 0$ (clearly, $\langle\nu^2|\infty\rangle=\sigma_\nu^2=2\sigma_u^2$).
Neglecting caustics would lead to $O(S)$ collision rates,
as predicted in the theory of Saffman and Turner \cite{saffman56}.
The multivalued velocity distribution in caustics appears to be 
crucial in producing high enough collision rates.

In the opposite limit $S\gg 1$, the particles are scattered by 
the velocity fluctuations they cross in their motion 
as if undergoing Brownian diffusion \cite{abrahamson75}.
In the $S\gg 1$ limit, the velocity difference equation can be approximated
by a stochastic differential equation (SDE) in the form,
choosing units such that $\tau_S=\sigma_u=1$:
\beq
\dot\nu_\alpha+\nu_\alpha=b_{\alpha\beta}(\r)\xi_\beta,
\qquad
\dot r_\alpha=\nu_\alpha,
\label{Langevin}
\eeq
where 
$b_{\alpha\gamma}(\r)b_{\gamma\beta}(\r)=(4/S)[g_{\alpha\beta}(0)-g_{\alpha\beta}(\r)]$,
$g_{\alpha\beta}(\r)=\sigma_u^{-2}\langle u_\alpha(\r,t)u_\beta(0,t)\rangle$, 
and $\xi_\alpha$ is the white noise: 
$\langle\xi_\alpha(t)\xi_\beta(0)\rangle=\delta_{\alpha\beta}\delta(t)$. In
the incompressible case: $\partial_\alpha g_{\alpha\beta}(\r)=0$.
For $S\gg 1$, the majority of particles pairs  at $r=0$, will have been at
$r\gg \rv$ for most of the previous time interval of length $\sim\tau_S$
of which they have memory, so that it is possible to set, in the first approximation:
$b(r)\simeq b(\infty)$. This leads to a Brownian collision dynamics, with a
velocity distribution of width $\sigma_\nu\sim S^{-1/2}\sigma_u$.

\vskip 10pt
Let us consider the effect of clusters on the collision velocity of the particle pairs,
in the small Stokes numer regime.
This regime could be analyzed within an SDE approach, 
imposing an artificially short correlation time to the random field
$\tau_E\ll \sigma_u/\rv$. This is a Kraichnan model regime \cite{kraichnan94},
in which the role of effective correlation time is played by the
diffusion time for a pair of tracers, $\tilde\tau_E=\rv^2/(\sigma_u^2\tau_E)$, 
to reach separation $\rv$. We have therefore an effective Stokes number
\beq
\epsilon=\tau_S/\tilde\tau_E=\tau_S\tau_E\sigma_u^2/\rv^2,
\label{epsilon}
\eeq
and it is possible to have a small-$\epsilon$ large-$S$ regime, in which
Eq. (\ref{Langevin}) continues to be valid. The small $\epsilon$ regime 
has been the subject of extensive study (see e.g.
\cite{wilkinson03,wilkinson05,duncan05}). For small $\epsilon$,
the particle phase is still monodisperse away from caustics, 
only with velocity not locally equal 
to that of the random field, 
as is instead
(to first approximation) 
in the $S\ll 1$, $\tau_E\sim \rv/\sigma_u$ regime. 

For small $\epsilon$, Eq. (\ref{Langevin}) leads 
to $\r$ changing little in a time $\tau_S$ (the correlation time for 
$\bnu$) and caustics arise as extreme events, 
in which a strong fluctuation in the random field causes particle pairs
to jump ballistically to zero separation in a time $\sim\tau_S$ 
\cite{wilkinson03,wilkinson05}.
Clusters affect the process
privileging particle pairs that are initially closer, i.e. less energetic fluctuations
in $\Delta_\r\u$, leading to smaller collision velocities \cite{note}. 
We give a
quantitative description of this effect for $D=1$ and 
$|r|\ll \rv$.

For $\epsilon\to 0$, the shapes of the pair trajectories terminating with 
a given collision velocity, will concentrate around the one that
maximizes probability (no condition is imposed on the caustics in which the
trajectories develop).
To determine the most likely trajectory ending with 
collision velocity $\nu(t)=\bar\nu$ at $t=0$, one can proceed
iteratively, from some initial guess for the separation history, 
say $r_0(t)=\bar\nu t$. (We assume immaterial particles, 
so that they can overlap without interaction).
For a smooth field, we can take for $|r|\ll\rv$ $g(r)\simeq 1-\alpha (r/\rv)^2$ with $\alpha=O(1)$,
which gives $b(r)\sim\epsilon^{1/2}r$.
For $|r|\ll \rv$, at the $k$-th step in the iteration procedure, 
the first of Eq. (\ref{Langevin}) will then read, apart of an $O(1)$ factor
in front of the right hand side (RHS): 
$\dot\nu_k+\nu_k=\epsilon^{1/2} r_k\xi$, leading to the solution
\beq
\nu_k(t)=\epsilon^{1/2}\int_{-\infty}^t\d\tau\,\ex^{\tau-t}r_k(\tau)\xi(\tau),
\label{solution}
\eeq 
where $\dot r_k=\nu_{k-1}$, $r_k(0)=0$, $k>0$.
The minimum problem will be, therefore,
\beq
\delta[\log{\cal P}[\xi]+\lambda_k\nu_k(0)]=0,
\label{variation}
\eeq
where $\delta$ indicates variation in $\xi$, ${\cal P}[\xi]$ is the 
functional PDF of the history $\xi(t)$ with $t\in [-\infty,0]$, $\lambda_k$
is the Lagrange multiplier to enforce $\nu_k(0)=\bar\nu$ and
the $r_k$ entering $\nu_k$ is assigned from the previous iteration.
From ${\cal P}[\xi]=\exp(-\frac{1}{2}\int\xi^2(t)\d t)$ and Eq. 
(\ref{solution}), the minimum problem Eq. (\ref{variation}) leads
to the history $\xi_k(t)=\lambda_k\epsilon^{1/2}\,\ex^t\,r_k(t)$.

Substituting $\xi_k$ into Eq. (\ref{solution}) 
and imposing $\nu(0)=\bar\nu$ gives
the velocity profile $\nu_k(t)$ in function of $r_k$, which, 
substituting into $\dot r_{k+1}=\nu_k$
leads to the iterative relation for $r_k(t)$:
\beq
\begin{array}{ll}
r_{k+1}(t)&=\bar\nu\{1-[\int_{-\infty}^0\d\tau\,\ex^{2\tau}r_k(\tau)]^{-1}
\\
&\times
[\int_{-\infty}^t\d\tau\,\ex^{2\tau-t}r_k^2(\tau)
+\int_t^0\d\tau\,\ex^\tau r_k^2(\tau)]\}
\end{array}
\label{iteration}
\eeq
The iterative scheme can be implemented numerically and converges rapidly.
In particular, the starting point of a jump terminating in a collision at 
velocity $\bar\nu$ is $r_{\bar\nu}\simeq\lim_{k\to\infty}r_k(-\infty)$
and one finds from Eq. (\ref{iteration}) $r_{\bar\nu}\simeq -7\,\bar\nu\tau_S$.
The rather large factor $7$ implies  that the jump does not start from a 
spatially localized ''kick'', rather, the sling acts over an 
$O(\bar\nu\tau_S)$ distance equivalent to that of the final free flight. 

From the same condition $\nu(0)=\bar\nu$, one finds
$\lambda_k=[\epsilon\int_{-\infty}^0\d\tau\,\ex^{2\tau}r_k(\tau)]^{-1}\bar\nu$,
which, substituted into $\xi_k(t)=\lambda_k\epsilon^{1/2}\,\ex^t\,r_k(t)$,
together with $r_k\sim\bar\nu$, gives for the noise history:
$\xi_k\sim\epsilon^{-1/2}$ and therefore for the PDF of observing a collision 
velocity $\bar\nu$ in a given pair will be, for $a\ll r_{\bar\nu}$:
$\rho_{jump}(\bar\nu|a)\simeq\rho_{jump}(\bar\nu|0)\sim {\cal P}[\xi_k]\sim\exp(-c/\epsilon)$,
with $c$ a constant that can be shown to be $1/6$ in 1D \cite{wilkinson03}.
The PDF of the collision velocity $\bar\nu$ in the given pair, generated by
a jump originating at separation $r$ will be instead:
$\rho_{jump}(\bar\nu|0)\delta(r-\bar\nu T)$ where $T\simeq 7\tau_S$. 

Multiplying by the PDF $\rho(r)$ of 
finding a pair at separation $r$, and integrating over $r$, gives 
the PDF of a collision at velocity $\bar\nu$
\beq
\rho_{coll}(\bar\nu)
=\rho(r_{\bar\nu})\rho_{jump},
\label{P_jump}
\eeq
with the cluster contribution contained in the PDF for the particle pair separation 
$\rho(r_{\bar\nu})$. A collision velocity $\bar\nu$ implies a permanence time 
$\propto|\bar\nu|^{-1}$ in an interval $\d r$, and this allows to write
the collision velocity PDF (that is the collision rate at that velocity) 
in the form
\beq
\rho_{coll}(\bar\nu)\sim\rho(\bar\nu,a)|\bar\nu|=\rho(a)\rho(\bar\nu|a)|\bar\nu|.
\label{rate1}
\eeq
This in turn can be substituted into Eq. (\ref{rate}) 
exploiting the relation, valid in $D$ generic: 
$\langle n(\r)n(0)\rangle=\Omega\bar n^2\rho(\r)$,
with $\Omega$ the domain volume and $\bar n$ the 
mean concentration. 

\begin{figure}
\begin{center}
\includegraphics[draft=false,width=6.5cm]{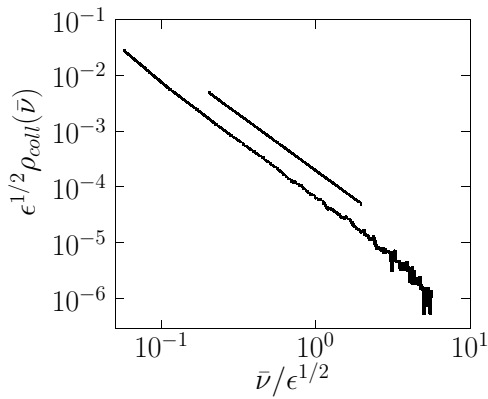}
\caption{
Collision velocity PDF for the 1D problem $\dot\nu+\nu=\epsilon^{1/2}(r+\alpha)\xi$,
$\dot r=\nu$, compared to $\rho=\nu^{-2}$ (upper line). Values of the parameters:
$\epsilon=0.04$
and $\alpha=0.02$
(a finite molecular diffusivity is necessary to regularize the dynamics at $r\to 0$).
}
\label{caufig1}
\end{center}
\end{figure}

In 1D, $\rho(r)\propto r^{-2}$ \cite{wilkinson03}, so
that $\rho_{coll}(\bar\nu)\propto\bar\nu^{-2}$, as confirmed in Fig \ref{caufig1}. 
In the absence of clustering, in comparison, the distribution would have been uniform
and $\sim \exp(-c/\epsilon)$ up to $|\bar\nu|\sim \rv/\tau_S$.
%
%
%
%
%

This picture extends to $D>1$, as the extremal trajectories are still straight lines,
and the relevant parameter remains the separation $r_{\bar\nu}$ at the start of the
jump. In this case the finite particle size must be taken into account and 
a jump ending in a collision will develop along a straight line that does not
necessarily pass through the center of the other particle.
The probability of a jump originating at $\r$ leading to collision will be proportional
therefore to the angle with which a particle is seen at distance $r$, i.e. $(a/r)^{D-1}$.
Nevertheless, if $a\ll r_{\bar\nu}$, the collision velocity PDF, i.e. the PDF for the
velocity in the direction of the jump at the other particle position, 
could be approximated as: $\rho_{jump}(\nu|a)\simeq\rho_{jump}(\nu|0)$.

\begin{figure}
\begin{center}
\includegraphics[draft=false,width=6.5cm]{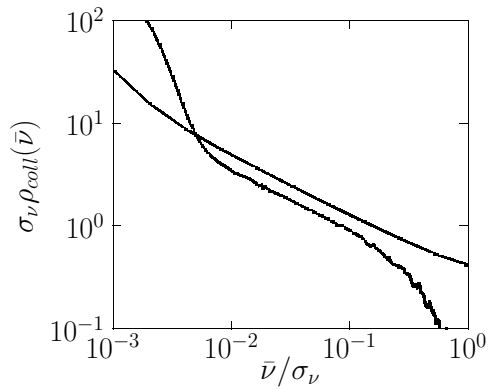}
\caption{
Collision velocity PDF in 2D from a sample of $10^5$ particles with $\epsilon=0.04$, 
advected by a smooth Kraichnan random field in a square domain $3\rv\times 3\rv$. 
The almost straight line is $\rho(\r)=r^{-2}\langle N_r^2\rangle$ 
(appropriately rescaled),
corresponding to a correlation dimension $D_2\simeq 1.45$.
}
\label{caufig2}
\end{center}
\end{figure}

Now, the number of particles in a shell at distance $r$ will be
$\propto\rho(\r)r^{D-1}\d r$, where the PDF $\rho(\r)$,
provided $\lim_{r\to 0}r^D\rho(\r)<\infty$, 
is related to the correlation dimension $D_2$ of the 
distribution by the equation
$\rho(\r)r^D\propto r^{-D}\langle N^2_r\rangle\propto r^{D_2}$,
with $N_r$ is the number of particles in a volume of linear size $r$.
Taking the product of the different contributions, the probability of collisions 
at velocity between $\bar\nu$ and $\bar\nu+\d\bar\nu$ will be therefore
\beq
\rho_{coll}(\bar\nu)\d\bar\nu\propto\rho_{jump}(\bar\nu|0)\rho(\r_{\bar\nu})
a^{D-1}\d r_{\bar\nu},
\label{3D}
\eeq
where $r_{\bar\nu}=\bar\nu T$ for some $T=O(\tau_S)$. 
If, as in the 1D case, $\rho_{jump}(\nu|0)$ is independent of  $\nu$, 
the following result will hold: 
\beq
\rho_{coll}(\bar\nu)\propto \bar\nu^{D_2-D}.
\label{scaling}
\eeq
Simulating the trajectories of an ensemble of inertial particles in a 2D Kraichnan 
random field leads to the result in Fig. \ref{caufig2}, which confirms the prediction
of Eq. (\ref{scaling}). 
The peak to the left is produced by the finite size of the particles,
in the present case $a\simeq 1/2056$ of the domain size.
Its width is $\langle\nu^2|a\rangle^{1/2}\sim a\sigma_\nu/\rv$ (the typical
relative velocity at separation $a$) and its height is
a factor $\sim\exp(c/\epsilon)$ above the scaling range to the right,
which is associated with the jumps.
%
%

\begin{figure}
\begin{center}
\includegraphics[draft=false,width=5.cm]{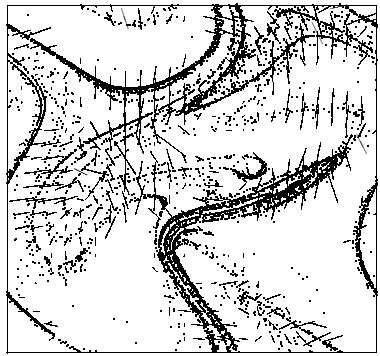}
\caption{
A snapshot of the particle distribution from the numerical simulation
of Fig. \ref{caufig2}, 
superinposed to the pattern of relative velocities (the small segments), which 
identify the caustics.
The length and direction of the segments correspond to the largest eigenvalue 
and associated eigenvector of the correlation matrix
$\langle \nu_i\nu_j|\x,t\rangle$, calculated as average on the particles 
in a box of size $a$ at $\x$.
}
\label{caufig3}
\end{center}
\end{figure}

As illustrated in Fig. \ref{caufig3}, the collisions actually take place
between clusters.
The scaling
$D_2<D$ causes closer clusters and therefore less energetic collisions
to be more likely.
Notice that, although clustering hinders high velocity collisions,
higher velocities are actually more probable inside clusters.
The reason is purely statistical: higher particle concentrations
are produced where clusters collide, i.e. by definition the place where
$\nu$ is larger. 
%
%
\vskip 10pt
For large Stokes numbers, collisions cease to occur as extreme events; hence,
the correspondence between collision velocities and jump lengths disappears,
and the predictions of Eqs.  (\ref{P_jump},\ref{scaling}) cease to be valid. 
In this limit, the particle velocity distribution ceases to be
monodisperse pointwise (near caustics, it would be a superposition of discrete 
jets) and particle velocities at close separations are in the first approximation 
independent. It was suggested in \cite{olla07} that concentration fluctuations 
are produced by slowly approaching particle pairs, which spend a significant 
time at separations $r\lesssim\rv$. To verify this, however, it is necessary
to separate out a cluster contribution in the velocity PDF $\rho(\bnu|\r)$, 
analogous to the one identified in  Eqs.  (\ref{rate1}-\ref{scaling}), or
equivalently in the relation
$\rho(\bar\nu,a)\propto |\bar\nu|^{-1}\rho_{jump}(\bar\nu|a)\rho(r_{\bar\nu})$.

Notice that the simultaneous increase of $\rho(r)$ and decrease in
$\langle\nu^2|r\rangle$ as $r\to 0$ is not sufficient to conclude that
concentration fluctuations are associated with smaller relative velocities;
in fact, $\lim_{r\to 0}\langle\nu^2|r\rangle= 0$ also for passive scalars in 
incompressible flows,
in which case, concentration fluctuations are absent. An alternative approach is
therefore required. 

As carried on in 
\cite{olla08}, a clustering part in $\rho(\bar\bnu,\bar\r)
=\langle\delta(\bnu(t)-\bar\bnu)\delta(\r(t)-\bar\r)\rangle$ 
could be identified in the higher order contributions from the
expansion of $\r(t)$ around the Brownian motion limit.
The analysis in \cite{olla08} indicates a strong dependence on compressibility of
the random field, and that the conclusion of the present paper, that clustering 
decreases collision velocities, remains valid at large $S$ only for compressible
flows. As in the small $\epsilon$ regime, however,
this clustering contribution cannot be identified with the
actual velocity distribution inside the clusters, and is more in the form of
a non-local cluster contribution to the velocity dynamics.

\end{document}